# Designing New Improper Ferroelectrics with a General Strategy

Ke Xu[1,2], Xue-Zeng Lu[3], and H. J. Xiang[1*]

[1]Key Laboratory of Computational Physical Sciences (Ministry of Education), State Key Laboratory of Surface Physics, Collaborative Innovation Center of Advanced Microstructures, and Department of Physics, Fudan University, Shanghai 200433, P. R. China

[2]Hubei Key Laboratory of Low Dimensional Optoelectronic Materials and Devices, Hubei University of Arts and Science, Xiangyang, 441053, P. R. China

[3]Department of Materials Science and Engineering, Northwestern University, Evanston, Illinois 60208, USA

Email: hxiang@fudan.edu.cn

**ABSTRACT**

The presence of a switchable spontaneous electric polarization makes ferroelectrics ideal candidates for the use in many applications such as memory and sensors devices. Since known ferroelectrics are rather limited, finding new ferroelectric (FE) materials has become a flourishing field. One promising route is to design the so-called hybrid improper ferroelectricity. However, the previous approach based on the Landau theory is not easily adopted to the systems that are unrelated to the *Pbnm* perovskite structure. To this end, we develop a general design rule that is applicable to any systems. By combining this rule with density functional theory calculations, we identify previously unrecognized classes of FE materials. It shows that the $R\bar{3}c$ perovskite structure can become FE by substituting half of the B-site cations. $ZnSrO_2$ with a non-perovskite layered structure can also be FE through the anion substitution. Moreover, our approach can be used to design new multiferroics as illustrated in the case of fluorine substituted $LaMnO_3$.

Recently ferroelectrics have attracted much attention due to their wide range of applications, especially in the electronic devices such as nonvolatile memory [1,2], tunable capacitors [3], solar cell [4], and tunnel junctions [5]. In the traditional proper ferroelectrics, such as cubic $BaTiO_3$, the transition metal ion $Ti^{4+}$ with $d^0$ configuration can hybridize with the oxygen 2p states leading to the FE phase transition due to the pseudo Jahn-Teller effect [6-8]. However, proper ferroelectrics are very few in nature [9]. To find more high-performance ferroelectrics, improper ferroelectrics becomes an intense research field [10] since the improper ferroelectricity was discovered in the artificial superlattice $PbTiO_3/SrTiO_3$. In this system, the ferroelectricity is induced by a trilinear coupling ($PQ_1Q_2$) between the FE mode (P) and two oxygen octahedral rotational modes ($Q_1$ and $Q_2$, respectively) [11]. Later, Benedek and Fennie predicted a different trilinear coupling existed in the double-layered Ruddlesden-Popper (RP) perovskite $Ca_3B_2O_7$ (B=Mn,Ti), which includes a FE mode, an oxygen octahedral rotational and an oxygen octahedral tilt modes [12], and named this type of improper ferroelectricity as "hybrid improper ferroelectricity" (HIF). It is worth noting that the HIF has been confirmed experimentally in $(Ca,Sr)_3Ti_2O_7$ single crystals [13], $Ca_3Ti_2O_7$ and $Ca_3Mn_2O_7$ polycrystalline crystals [14] and $Ca_3(Ti,Mn)_2O_7$ ceramics very recently [15]. The HIF was also found in the 1:1 A-cation ordering perovskite-type superlattice [16-19], A-cation ordering RP $NaRTiO_4$ (R=Y,La,Nd,Sm-Ho) [20], B-cation ordering $(YFeO_3)_2/(YTiO_3)_2$ superlattice [21], and a metal-organic perovskite material [22].

In the above-mentioned works, one usually starts from a high-symmetry structure (e.g., cubic perovskite structure), then the effect of atomic substitution and soft phonon modulations are examined to see whether the ferroelectricity can be induced. Finally, the trilinear coupling mechanism is discussed to understand the origin of the improper ferroelectricity. This procedure is informative. However, it is tedious and its applicability to other type of compounds is limited since even the high symmetry structure may be unknown. Therefore, two key problems remain to be addressed: Is it possible to propose a general method to design improper ferroelectrics? Can improper ferroelectricity be obtained in systems with totally different structures from perovskite? The answers to these questions may widen the opportunities of finding the improper FE materials. In this work, we propose a general approach to design the improper FE materials. Our approach can not only reproduce the previous results (e.g., 1:1 A-cation [001] ordering FE perovskite superlattices) but also predict new improper ferroelectrics. Our results show that the B-cation order [11$\bar{1}$]-superlattice La$_2$(Co,Al)O$_6$ with a $R\bar{3}c$ parent structure displays a spontaneous polarization. And ZnSrO$_2$ with a non-perovskite layered structure can also be FE through the anion substitution. Moreover, the LaMnO$_3$ with the fluorine substitution can become multiferroics with a sizable polarization with its direction perpendicular to the direction of the magnetization. Hence, our study may pave a new route to find the improper FE materials.

The geometry optimization and electron structure calculation are performed by the state-of-the-art density functional theory (DFT) [23] using projector augmented-wave (PAW) [24] potentials as implemented in the Vienna ab-initio Simulation Package (VASP) [25,26]. In this DFT plane-wave calculations, the plane wave cutoff energy is set to 500 eV, and the exchange-correlation interactions are described by the Perdew–Burke–Ernzerhof (PBE) generalized gradient approximation (GGA) [27]. In order to take into account for the proper orbital dependence of the onsite Coulomb and exchange interaction, we employ the GGA + U method [28] in treating the Co's and Mn's 3d orbitals and the U are set to 7.0 eV and 3.0 eV, respectively. For relaxation of the structure, the Hellmann-Feynman forces on each atoms are less than 0.001 eV Å$^{-1}$. The total electric polarization is calculated using the Berry phase method [29,30]. We adopt a global optimization method based on the genetic algorithm [31] to search the ground state of the layered La$_2$CoAlO$_6$. In our calculation, the basic lattice structure is fixed and our purpose is to find the distortions that leads to the lowest energy. In this GA searching method, DFT calculation is adopted to relax the structure and the low-spin configuration is adopted for Co. The number of atoms in the supercell is fixed to 20. The population size and number of generations are set to 16 and 10, respectively.

**General design guidelines:** Firstly, the parent structure should have the true inversion centers and pseudo inversion centers simultaneously and also a sizable band gap. The true inversion center (TIC) is defined as a position on which there is inversion symmetry *I* of the structure. The pseudo inversion center (PIC) is a position which can become a TIC after small displacements of the atoms. With a given centrosymmetric structure (i.e., parent structure), we first find all the TICs and PICs of the parent structure. Then we find out all the possible atom substitutions within a given supercell of the parent structure, which lift all the TICs but still keep at least one PIC. These atom substituted structures are candidates for ferroelectrics. After the slight displacements of the ions of the element substituted structure so that the PIC becomes a TIC, the corresponding paraelectric (PE) structures can be obtained. Fig. 1(a) shows a chart for the general strategy for designing new ferroelectrics. Our method is illustrated clearly with a simple one-dimensional toy model [see Fig. 1(b)]. Assuming that there is a one-dimensional chain in which there are two atoms of the same type (represented by A) in a unit cell and the two atoms locate at 0.0 and 0.4, respectively. Then, one can find two TICs denoted by the crosses within a unit cell before substitution. If replacing an A atom by a B atom in the unit cell, all the TICs are broken leading to a possible polar structure with P along the left or right directions. However, there is a PIC at 0.5 even after the substitution. When moving the B atom to the center of the unit cell of the lattice, the PIC will become a TIC. From the above discussion, the key point in our method is lifting all the TICs by atom substitution and maintaining at least one PIC in the system. To ensure the successful design of the ferroelectrics, we need three additional conditions: (I) The FE phase should be locally stable; (II) Both FE and PE phases should have large band gaps.

**Improper ferroelectrics based on the *Pbnm* ABO$_3$ perovskite structure:** Let us first concentrate on the design of the ferroelectrics based on ABO$_3$ perovskites. The *Pbnm* structure with 20 atoms in the unit cell can be obtained if the cubic *Pm$\bar{3}$m* structure undergoes an out-phase rotation and an in-phase rotation of oxygen octahedral ($a^-a^-c^0$ and $a^0a^0c^+$ in Glazer notation, respectively) [9]. By using our method, we find that there are 8 TICs (all B-sites are TICs) and 24 PICs in one unit cell of the *Pbnm* structure (see Fig. S1 for details in the supplementary materials [32]). Within the 20-atom cell, it is found that there are 14 possible substitution-induced ferroelectrics in total, among which there are 2 types of A-site substitution and 12 types of anion substitution (see Fig S2). One of the A-site substitution $(A/A')B_2O_6$ is the [001] FE superlattice discovered in previous studies [16-18]. The A and A'-sites ordering breaks all the TICs but maintain the 16 PICs. The net electric polarization is due to the fact that A-site displacement and A'-site displacement are not complete cancelled. We find that it is impossible to obtain improper ferroelectrics through B-site substitution in a 20-atom cell, in agreement with the previous result [18]. But when considering a twofold supercell along the z direction, we can obtain ferroelectrics through B-site substitution. For example, our result indicates that the B/B' cation ordered 2/2 supercell adopts the FE $Pmn2_1$ symmetry (see Fig. S3 for the possible substitutions in the 2/2 B/B' cation ordering superlattices). All these results demonstrate that our method can not only reproduce the previous results [16-18], but also predict previously unknown FE materials, particularly, we point out for first time that the anion substitution may lead to improper ferroelectrics, as will be discussed later in detail.

**Improper ferroelectrics based on the $R\bar{3}c$ $ABO_3$ perovskite structure:** Another family of perovskite oxides take the 10-atom $R\bar{3}c$ structure as the lowest energy structure [33], which has a rhombohedral $a^-a^-a^-$ tilt pattern around the [111] direction with respect to $Pm\bar{3}m$ structure. There are 8 TICs and 8 PICs in one unit cell (the B-sites are TICs). Within the 10-atom cell, we find that there are two possible ways of the anion substitution to induce improper FE [see Fig S4 (a) and (b)]. If we consider a twofold 20-atom supercell, there are 3 types ferroelectrics which result from the A-site, B-site and C-site substitutions, respectively [see Fig S5 (a), (b) and (c) for A-site, B-site and C-site substitutions, respectively]. Note that the A-site order in the 1:1 superlattice exhibits a non-FE structure with the $R32$ symmetry [34].

In the following, $LaCoO_3$ and $LaAlO_3$ are selected as the parent structures to demonstrate the B-site substitution induced ferroelectricity in the $R\bar{3}c$ $ABO_3$ perovskite. The mismatch in the lattice constants between these two compounds is only about 0.13%. The $LaCoO_3$ with $R\bar{3}c$ symmetry adopts a low-spin nonmagnetic ground state ($t_{2g}^6$, S = 0) for $Co^{3+}$ [35-37]. For all the perovskites with $R\bar{3}c$ symmetry, we find the [$11\bar{1}$] B-site order in the 1:1 20-atom superlattice results in a polar structure with $C2$ symmetry. The corresponding PE structure takes $R\bar{3}m$ symmetry. Both FE and PE structures have large band gaps about 2.0 eV, and the energy barrier between FE and PE structures is ~0.3 eV for a 20-atom cell. This spontaneous electric polarization is calculated to be 0.5 μC/cm$^2$, which is along the [1$\bar{1}$0] direction. The ferroelectricity is caused by the fact that the B-site inversion symmetry are broken by the B-site substitution. For clarity, we show the local structures of LaAlO$_3$ and La$_2$CoAlO$_6$ to understand the direction of polarization. A 10-atom La$_3$Al$_4$O$_3$ cluster in $R\bar{3}c$ LaAlO$_3$ forms a tetrahedral structure. This cluster has one out-of-plane threefold-rotational axis on the central Al atom and three in-plane twofold-rotational axes along the La-La bond directions [see Fig 2(a)]. In La$_2$CoAlO$_6$ with a [11$\bar{1}$] B-site order, only one twofold rotation axis is kept due to the Co-atom substitution. This twofold rotation axis is exactly along the direction of the polarization [see Fig 2(b)]. Interestingly, we find that the direction of polarization **P** [1$\bar{1}$0] is parallel to the cross product of the B-site ordering vector **D** [11$\bar{1}$] and the octahedral rotational vector **Ω** [111] [see Fig 2(c)], i.e., $\boldsymbol{P} \parallel \boldsymbol{D} \times \boldsymbol{\Omega}$. This relation also holds for the other FE domains. The phonon dispersion also shows that the superlattice is dynamically stable [see Fig S7]. In addition, our genetic algorithm (GA) [31] structure search confirms that the FE $C2$ structure is indeed the lowest energy structure.

In proper ferroelectrics, the FE mode is the primary order parameter and there is a double well potential in the plot of energy versus polar displacement. But for the improper ferroelectrics, the FE mode is no longer the primary order parameter, i.e., the polarization is induced by one or two rotational modes [9,11,38]. To verify whether the [11$\bar{1}$]-superlattice LaCoO$_3$/LaAlO$_3$ is the improper ferroelectrics or not, the stability of the FE mode will be examined. We adopt the ISOTROPY software [39] to obtain the symmetry-adapted phonon modes in the low symmetry FE structure. To find out the appropriate FE mode and eliminate the effect of the symmetry breaking solely due to atomic substitution, we replace the Co atoms in La$_2$(Co,Al)O$_6$ back to Al atoms before doing the mode decomposition with respect to the cubic $Pm\bar{3}m$ LaAlO$_3$. We find two dominant modes, namely, ferroelectric $\Gamma_4^-$ mode and rotation $R_5^-$ mode, respectively. When applying the FE mode to the PE La$_2$(Co,Al)O$_6$ structure, we find the total energy increases with the mode magnitude, and thus the FE mode is not a soft mode. While for the rotational mode, it has a double well potential in the plot of energy versus rotational displacement [see Fig 2(d)]. These characteristics indicate that the layered (LaAlO$_3$)$_1$/(LaCoO$_3$)$_1$ superlattice is an improper ferroelectric.

**New ferroelectrics based on the non-perovskite structures:** Previous works on designing improper ferroelectrics mainly focused on the perovskite-related structures. We note that if a crystal structure has the TIC, PIC and band gap simultaneously, it is a possible candidate for designing new ferroelectrics through atom substitution. Here, we find that the non-perovskite structure ZnSrO$_2$ with *Pnma* symmetry is suitable for ferroelectrics design. The parent structure with 8 PICs and 8 PICs has a large direct band gap ~2.2 eV at the PBE level. The unit cell can be regarded as a layered structure along the z direction that contains two formula units of Zn$_2$Sr$_2$O$_4$, in which there exists three twofold screw axis 2$_1$ as shown in Fig. 3(a). By replacing the outer O atoms of the upper ZnSrO$_2$ layer by S atoms, we obtain a FE *Pmc*2$_1$ structure of Zn$_4$Sr$_4$O$_6$S$_2$. The corresponding PE structure has the *Pmma* symmetry. The energy barrier between FE and PE is ~0.4 eV. The replacement of the oxygen atoms with the two sulfur atoms in a unit cell will break the screw axes along the **a** and **c** directions and induce a large spontaneous polarization (~20.0 μC/cm$^2$) along the **b** direction [see Fig 3(b)]. The large polarization is due to the large atomic displacements induced by the sulfur replacement. Fig 3(b) shows the atomic direction and magnitude of the displacement (see red arrows) after the full relaxation. The S atom is nearly at the original O position. Because the bond length of Zn-S is larger than that of Zn-O, Zn$_1$ and Zn$_2$ ions move along the **-c** or **c** directions, respectively. This subsequently leads to a large displacement of the O$_2$ atom along the direction (i.e., approximately [011] direction) perpendicular to the Zn$_1$-Zn$_2$ bond direction. Overall, these ion displacement induce a large polarization along the -**b** axis in the relaxed structure of Zn$_4$Sr$_4$O$_6$S$_2$. Moreover, the new ferroelectrics Zn$_4$Sr$_4$O$_6$S$_2$ is an environmentally friendly non-toxic material unlike PbTiO$_3$.

**Multiferroics based on the anion-substituted perovskite compound:** Our method is also suitable for designing new multiferroics. LaMnO$_3$ with the *Pbnm* perovskite structure is selected as a parent structure. One can obtain a FE structure by replacing one-third of oxygen atoms by fluorine atoms to form a superlattice along the orthorhombic **a** direction [see Fig. 4 (a) and (b)]. We note that the replacement of oxygen ions in perovskite oxides by other anions such as fluorine ions has been achieved experimentally [40-42]. Since La and Mn ions in the parent $LaMnO_3$ structure are both trivalent, we also replace all the A-site La ions with a bivalent ion Sr, whose radius is very similar to that of $La^{3+}$. The chemical formula of the final substituted structure becomes $SrMnO_2F$. The FE and PE $SrMnO_2F$ phases take polar $Pmc2_1$ and non-polar $Pbcm$ symmetry, respectively. The detailed mechanism for the anion-substitution induced ferroelectricity is shown in Fig. S6. The electric polarization of the FE structure is 14.5 $\mu C \cdot cm^{-2}$ along the orthorhombic **b** direction. We find that the magnetic ground states for both FE and PE structures of $SrMnO_2F$ are the A-type antiferromagnetic (A-AFM), which is similar to the parent structure $LaMnO_3$ [43]. When considering the spin-orbit coupling (SOC) effect, $SrMnO_2F$ displays a magnetic anisotropy with the magnetic easy axis along the orthorhombic **b** direction. The Dzyaloshinskii-Moriya (DM) interaction [44,45] leads to a canted ferromagnetic magnetic moment of about 0.058 $\mu B$ along the orthorhombic **c** direction. In the parent structure $LaMnO_3$, the totally canted magnetic moment is 0.027$\mu B$, which indicates the anion substitution not only induces the ferroelectricity, but also increases the canted magnetic moment. Since the $SrMnO_2F$ with the $Pmc2_1$ structure has simultaneous weak ferromagnetism and ferroelectricity, it is a multiferroics. In addition, we find that the $LaMnO_2F$ with the FE $Pmc2_1$ structure has a G-type antiferromagnetic order with a very small magnetic anisotropy since the $Mn^{2+}$ ion is half-filled.

**Summary**

In conclusion, we proposed a general method to design new ferroelectrics. With this method, we can not only rediscover the hybrid improper ferroelectricity in A-site [001] ordering $Pbnm$ perovskite oxides, but also discover new improper ferroelectrics. In particular, our results show the B-site ordering [11$\bar{1}$] superlattice $(LaCoO_3)_1/(LaAlO_3)_1$ can be an improper ferroelectric. We also demonstrated for the first time that the anion substitution can be adopted to generate the ferroelectrics as exemplified in the case of perovskite systems and $ZnSrO_2$ with a non-perovskite structure. Moreover, through the anion substitution in $LaMnO_3$, we find that $SrMnO_2F$ is multiferroics, which indicates that our method is useful for the design of the multiferroic materials.

the $R\bar{3}c$ perovskite structure and other supplementary materials, which includes Ref [46].

**Acknowledgements**

Work at Fudan was supported by NSFC, FANEDD, NCET-10-0351, Research Program of Shanghai Municipality and MOE, the Special Funds for Major State Basic Research, Program for Professor of Special Appointment (Eastern Scholar), and Fok Ying Tung Education Foundation. K. X. was partially supported by NSFC 11404109. XL was supported in part by the National Science Foundation (NSF) through the Pennsylvania State University MRSEC under award number DMR-1420620. We thank Panshuo Wang for useful comments on the manuscript.

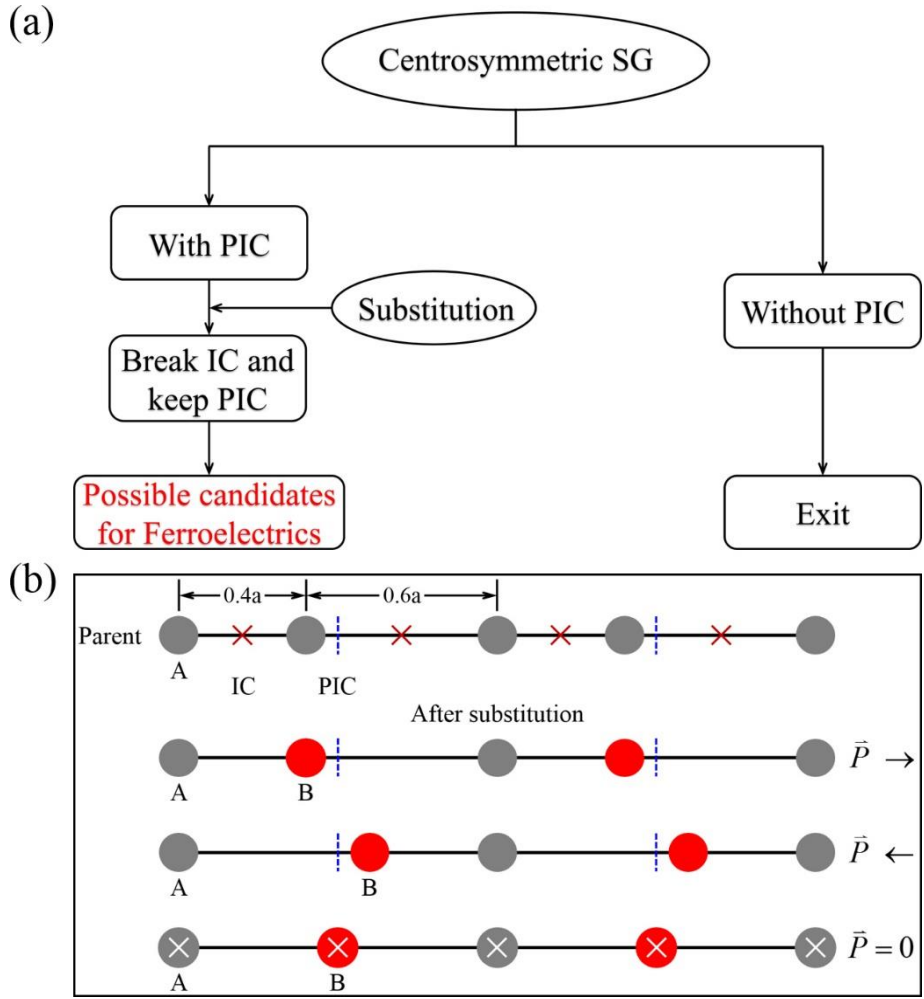

Fig 1. (a) The flow chart of our method for designing new ferroelectrics. (b) A simple 1D model system to illustrate the idea of our method. The position 0.5a is the pseudo inversion centers (PICs) and denoted by the vertical dashed line. The true inversion centers (TICs) are denoted by "×". "SG" is an abbreviation for "space group".

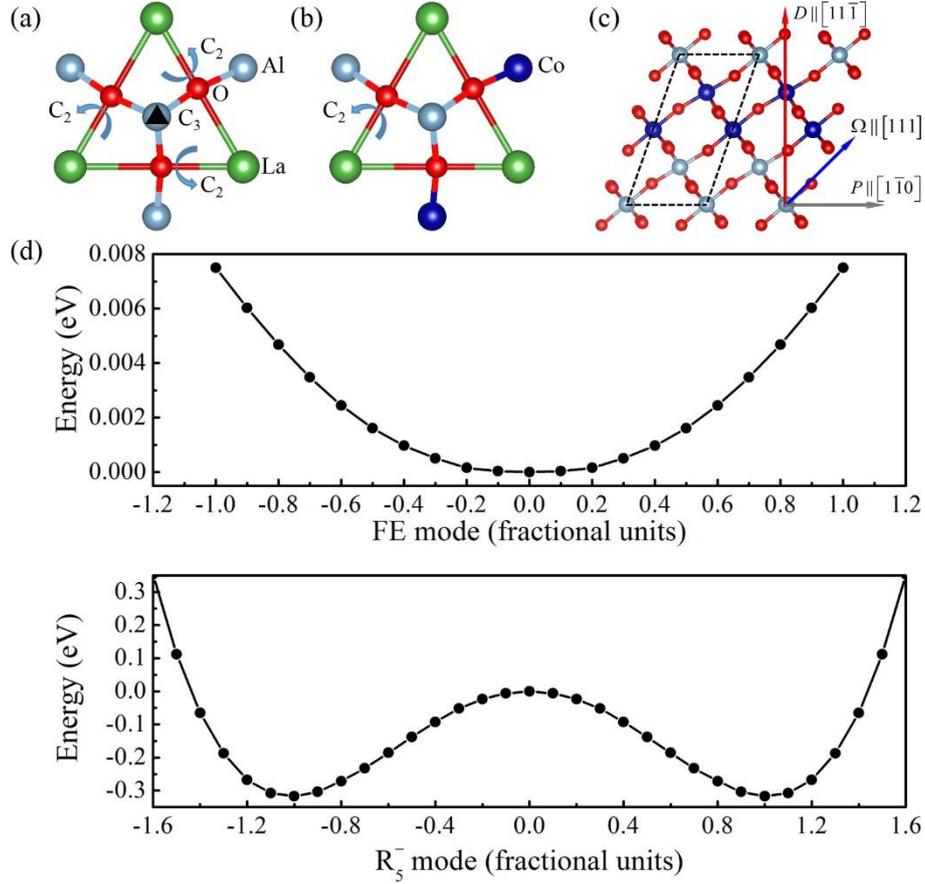

Fig 2. B-site substitution induced ferroelectricity in $R\bar{3}c$ perovskite system. (a) shows the structure of the La$_3$Al$_4$O$_3$ cluster in $R\bar{3}c$ LaAlO$_3$. The corresponding structure of the La$_3$Co$_2$Al$_2$O$_3$ cluster in $C2$ La$_2$(Co,Al)O$_6$ is displayed in (b). (c) The structure of the FE $C2$ La$_2$(Co,Al)O$_6$. The A-site La atoms are not shown for clarity. The direction of polarization **P** is perpendicular to both the superlattice direction **D** and the rotation direction **Ω** of the oxygen octahedron. (d) shows the total energies of La$_2$(Co,Al)O$_6$ as a function of the magnitudes of the FE mode and the rotation mode, respectively.

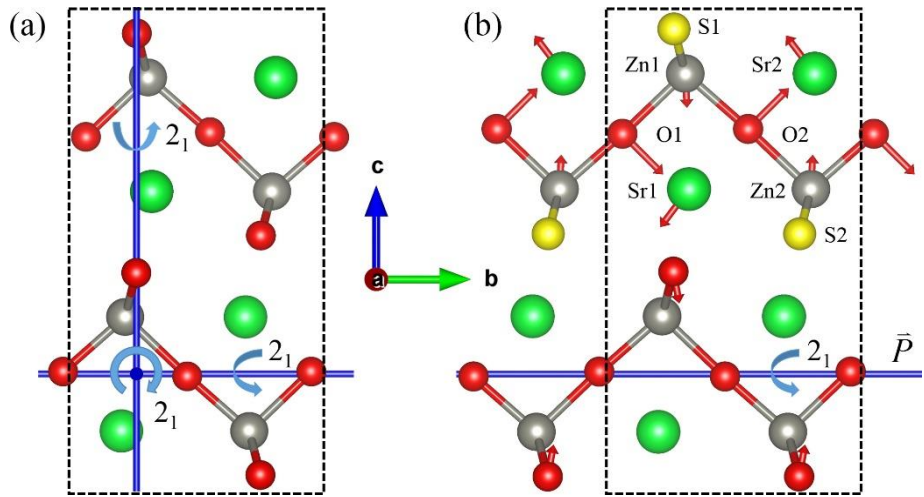

Fig 3. Anion substitution induced ferroelectricity in $ZnSrO_2$ with a non-perovskite structure. (a) The original structure of $ZnSrO_2$. The three screw axes are shown. (b) The FE structure of $Zn_4Sr_4O_6S_2$ obtained by replacing one fourth of the oxygen atoms by sulfur atoms. After the anion substitution, only the screw axis along the **b** axis is kept, resulting in a polarization along the **b** axis. The atomic displacements after relaxation are denoted by red arrows.

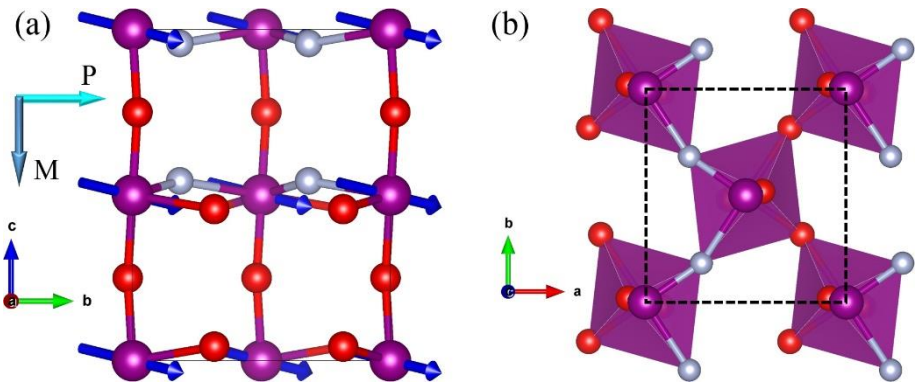

Fig 4. Multiferroic SrMnO$_2$F obtained by replacing one third of O ions by F ions in *Pbnm* LaMnO$_3$ (La ions are also replace by Sr ions to maintain the number of valence electrons). (a) The geometry and magnetic structure of SrMnO$_2$F. The Mn$^{3+}$ magnetic moments are denoted by blue arrows. Note that the canting of the magnetic moments along the **c** axis is exaggerated to guide the eye. The directions of the total electric polarization (**P**) and total canted moment (**M**) are depicted. (b) A polyhedral representation of the SrMnO$_2$F structure viewed along the orthorhombic **c** direction, which clearly shows the superlattice ordering of F and O ions along the orthorhombic **a** direction. The A-site cations are not shown for clarity.